\documentclass{article}
\usepackage{amsfonts}
\usepackage{amsmath}

\input{tcilatex}
\begin{document}

\title{Generalized Einstein gravities and generalized AdS symmetries}
\author{L. C\'{a}rdenas$^{1}$, J. D\'{\i}az$^{2}$, P. Salgado$^{2}$ and D.
Salgado$^{1}$ \\
$^{1}$Departamento de F\'{\i}sica, Universidad de Concepci\'{o}n, \\
Casilla 160-C, Concepci\'{o}n, Chile\\
$^{2}$Instituto de Ciencias Exactas y Naturales, Facultad de Ciencias,\\
Univesidad Arturo Prat, Avda. Arturo Prat 2120, Iquique, Chile.}
\maketitle

\begin{abstract}
We consider the curvatures $2$-form asociated with $AdS\mathcal{L}_{4}$%
-valued one-form gauge connetion, and then we construct a four-dimensional
action that generalize the Einstein-Hilbert gravity. It is shown that the
Maxwell extension of Einstein gravity can be obtained from $AdS\mathcal{L}%
_{4}$-gravity making use of the In\"{o}n\"{u}-Wigner contraction method. \
In the same way, by gauging the $AdS\mathcal{L}_{5}$-spacetime algebra, the
Einstein-Hilbert gravity is extended including the vector fields $k_{\mu
}^{ab}$ and $h_{\mu }^{a}$ which are associated with non-Abelian tensors and
non Abelian vectors charges in the $AdS\mathcal{L}_{5}$ algebra.

The $\mathfrak{B}_{5}$ extension of Einstein gravity can be obtained from $%
AdS\mathcal{L}_{5}$-gravity using of the above mentioned contraction
procedure. Some aspects of a gravity based on the algebra $AdS\mathcal{L}%
_{6} $ are considered in an appendix.
\end{abstract}

\section{\textbf{Introducci\'{o}n }}

Some time ago was extended the standard geometric framework of Einstein
gravity by gauging the Maxwell algebra \cite{azcarr}, \cite{azcarr1} which
led to a generalized cosmological term that includes a contribution from the
six four-vector fields $k_{\mu }^{ab}$ which introduce a new set of
curvatures denoted by $F_{\mu \nu }^{ab}$ (in addition to the well-known
torsion and Lorentz curvature) that allow to build generalizations of the
Einstein action.

The Maxwell algebra \cite{maxw1}, \cite{maxw2} can also be obtained from
(A)dS algebra using the Lie algebra expansion procedure developed in Refs. 
\cite{exp1}, \cite{exp2}, \cite{exp3}, \cite{bhv}, \cite{exp4}. This
procedure allows obtaining two families of algebras, which are known as
generalized Poincar\'{e} algebras (also called $\mathfrak{B}_{n}$ algebra$,$
where Maxwell's algebra corresponds to the particular case known as $%
\mathfrak{B}_{4}$ algebra\ ) and generalized AdS algebras (also called AdS$%
\mathcal{L}_{n}$ algebras) \cite{ss}, \cite{sor1}, \cite{sor3}.

An Expansion is, in general, an algebra dimension-changing process, i.e., is
a way to obtain new algebras of increasingly higher dimensions from a given
one. \ A physical motivation for increasing the dimension of Lie algebras is
that increasing the number of generators of an algebra is a non-trivial way
of enlarging spacetime symmetries. Examples of this can be found in \cite%
{azcarr}, \cite{azcarr1}, \cite{bonanos}, \cite{gomis}.

The generalized Poincar\'{e} algebra $\mathfrak{B}_{n}$ \cite{chs5}, \cite%
{ss}, \cite{chs1}, \cite{luk} can be obtained from anti-de-Sitter algebra
and the semigroup $S_{E}^{2n-1}=\left\{ \lambda _{0},\cdot \cdot \cdot
,\lambda _{2n}\right\} $ whose multiplication law is given by $\lambda
_{\alpha }\lambda _{\beta }=\lambda _{\alpha +\beta }$ when $\alpha +\beta
\leq 2n$ and $\lambda _{\alpha }\lambda _{\beta }=\lambda _{2n}$ when $%
\alpha +\beta >2n$, where $\lambda _{2n}$ corresponds to the zero element of
the semigroup. The generators of $\mathfrak{B}_{n}$ denoted by $\left(
P_{a},J_{ab},Z_{ab}^{(i)},Z_{a}^{(i)}\right) $ satisfy the following
commutation relations 
\begin{align}
\left[ P_{a},P_{b}\right] & =Z_{ab}^{\left( 1\right) },\text{ \ \ }\left[
J_{ab},P_{c}\right] =\eta _{bc}P_{a}-\eta _{ac}P_{b},  \notag \\
\left[ J_{ab},J_{cd}\right] & =\eta _{bc}J_{ad}+\eta _{ad}J_{bc}-\eta
_{ac}J_{bd}-\eta _{bd}J_{ac},  \notag \\
\left[ J_{ab},Z_{c}^{(i)}\right] & =\eta _{bc}Z_{a}^{(i)}-\eta
_{ac}Z_{b}^{(i)},  \notag \\
\left[ Z_{ab}^{\left( i\right) },P_{c}\right] & =\eta _{bc}Z_{a}^{(i)}-\eta
_{ac}Z_{b}^{(i)},\text{ }  \notag \\
\left[ Z_{ab}^{\left( i\right) },Z_{c}^{(j)}\right] & =\eta
_{bc}Z_{a}^{(i+j)}-\eta _{ac}Z_{b}^{(i+j)},  \notag \\
\left[ J_{ab},Z_{cd}^{\left( i\right) }\right] & =\eta _{bc}Z_{ad}^{\left(
i\right) }+\eta _{ad}Z_{bc}^{\left( i\right) }-\eta _{ac}Z_{bd}^{\left(
i\right) }-\eta _{bd}Z_{ac}^{\left( i\right) },  \notag \\
\left[ Z_{ab}^{\left( i\right) },Z_{cd}^{\left( j\right) }\right] & =\eta
_{bc}Z_{ad}^{\left( i+j\right) }+\eta _{ad}Z_{bc}^{\left( i+j\right) }-\eta
_{ac}Z_{bd}^{\left( i+j\right) }-\eta _{bd}Z_{ac}^{\left( i+j\right) }, 
\notag \\
\text{\ }\left[ P_{a},Z_{b}^{(i)}\right] & =Z_{ab}^{\left( i+1\right) }, 
\notag \\
\left[ Z_{a}^{(i)},Z_{b}^{(j)}\right] & =Z_{ab}^{\left( i+j+1\right) }
\label{ej5}
\end{align}%
where, $\tilde{J}_{ab}$ and $\tilde{P}_{a}$ are the generators of the
anti-de-Sitter algebra and $J_{ab}=\lambda _{0}\otimes \tilde{J}_{ab},$ $%
Z_{ab}^{\left( i\right) }=\lambda _{2i}\otimes \tilde{J}_{ab},$ $%
P_{a}=\lambda _{1}\otimes \tilde{P}_{a}$ y $Z_{a}^{(i)}=\lambda
_{2i+1}\otimes \tilde{P}_{a}$, with $i,j=0,1,\cdot \cdot \cdot ,n-1$.

From the commutation relations (\ref{ej5}) we see that the set $\mathfrak{B}%
_{n}^{I}=\left( P_{a},Z_{ab}^{(i)},Z_{a}^{(i)}\right) $ satisfies the
conditions $\left[ \mathfrak{B}_{n}^{I},\mathfrak{B}_{n}^{I}\right] \subset 
\mathfrak{B}_{n}^{I}$, $\left[ so(3,1),\mathfrak{B}_{n}^{I}\right] \subset 
\mathfrak{B}_{n}^{I}$, i.e. $\mathfrak{B}_{n}^{I}$ is an ideal of the
generalized Poincar\'{e} Algebra $\mathfrak{B}_{n}$ generated by $\left(
P_{a},Z_{ab}^{(i)},Z_{a}^{(i)}\right) $. This means that the generalized
Poincar\'{e} Algebra $\mathfrak{B}_{n}$ is the semidirect sum of the Lorentz
algebra $so(3,1)$ and the ideal $\mathfrak{B}_{n}^{I}$, that is, $\mathfrak{B%
}_{n}=so(3,1)\uplus \mathfrak{B}_{n}^{I}$.

From (\ref{ej5}) it is also possible to see that for $i=0,$ $n=1$ we have
the Poincare algebra that in this nomenclature we will call $\mathfrak{B}%
_{1} $ (also called $\mathfrak{B}_{3}$); for $i=1,$ $n=2$ we have Maxwell's
algebra which in this nomenclature we will call $\mathfrak{B}_{2}$ (also
called $\mathfrak{B}_{4}$); for $i=2,$ $n=3$ we have the algebra, $\mathfrak{%
B}_{3}$ (also called $\mathfrak{B}_{5}$); for $i=3,$ $n=4$ we have the
algebra $\mathfrak{B}_{4}$ (also called $\mathfrak{B}_{6}$), etc. These
algebras and their relation to AdS$\mathcal{L}_{n}$ algebras are shown in
Appendix A.

On the other hand the AdS$\mathcal{L}_{n}$ algebras \cite{ss}, \cite{sor1}, 
\cite{sor3} can be obtained from the anti-de-Sitter algebra and the
semigroup $S_{\mathcal{M}}^{(N)}=\left\{ \lambda _{\alpha }\right\} _{\alpha
=0}^{N}$ whose multiplication law is given by $\lambda _{\alpha }\lambda
_{\beta }=\lambda _{\alpha +\beta }$ when $\alpha +\beta \leq N$ and $%
\lambda _{\alpha }\lambda _{\beta }=\lambda _{\alpha +\beta -2\left[ (N+1)/2%
\right] }$ $\alpha +\beta >N$, where $\left[ x\right] $ is the integer part
of $x$. Note that for $N$ odd the semigroup $S_{\mathcal{M}}^{(N)}$
coincides with the cyclic group of $\left( N+1\right) $ elements $%
\mathbb{Z}
_{N+1}$.

It might be of interest to mention that for the AdS$\mathcal{L}_{n}$
algebras with $n=3,4,5,6$ it is found that in the cases $n=3$ and $5$ they
cannot be expressed as a semidirect sum of the Lorentz algebra and an ideal,
while for the cases $n=4$ and $6$ it is straightforward to see that they can
be written as the semidirect sum of the Lorentz algebra and an ideal (see
Appendix A).

The consequences of considering a space-time with Maxwell symmetries in the
description of the gravitational field have been studied, in the context of
Chern-Simons gravities, in Refs. \cite{ss}, \cite{chs1}, \cite{luk}, and in
the context of four-dimensional gravity, in Refs. \cite{azcarr}, \cite%
{azcarr1}.

The implications of the existence of space-times with symmetries given both,
by generalized Poincar\'{e} algebras and by generalized AdS algebras, were
studied, in the case of odd dimensions, in references \cite{chs3}, \cite%
{chs4}, \cite{chs5}, and in the even dimensions, in the context of WZW
terms, in Refs. \cite{even1}, \cite{even2}, \cite{even3}, \cite{even4}.

An interesting open problem is to obtain an understanding of the
gravitational field from the direct construction of $4$-dimensional
invariant actions both under the generalized Poincare algebras and the AdS$%
\mathcal{L}_{n}$ algebras.

The geometry of space-times based on generalized Poincare algebras and
generalized AdS-Lorentz algebras involve new gauge fields and therefore new
tensor curvatures that allow to construct new gravity actions which lead to
modifications of the standard gravity.

This paper is organized as follows: In Section $2$ we consider the
construction of the curvatures $2$-form asociated with AdS$\mathcal{L}_{4}$%
-valued one-form gauge connetion, and a four-dimensional gravity based in
the AdS$\mathcal{L}_{4}$ algebra is constructed using the procedure
described in references \cite{azcarr}, \cite{azcarr1}. In Section $3$ we
construct a four-dimensional action, using an AdS$\mathcal{L}_{5}$-valued
one-form gauge connection following the same procedure of Section $2$.
Actions based on the generalized Poincare algebras ($\mathfrak{B}_{4}$ and $%
\mathfrak{B}_{5}$) are obtained in Section $4$ by making use of the
well-known contraction methods. Finally Concluding Remarks are presented in
Section $5$. \ Two Appendices are included, where are considered details
about the algebras $\mathfrak{B}_{n}$ and AdS$\mathcal{L}_{m}$ as well as
some aspects of the construction of an action gravity based in an AdS$%
\mathcal{L}_{6}$-valued one-form gauge connetion$.$

\section{\textbf{AdS}$\mathcal{L}_{4}$\textbf{\ extension of
Einstein-Hilbert-Cartan\ gravity}}

In this section we will study the extension of Einstein's theory of
gravitation considering AdS$\mathcal{L}_{4}$ symmetries.

\subsection{\textbf{Gauging the }\textbf{AdS}$\mathcal{L}_{4}$\textbf{%
-algebra}}

In order to write down the two-form curvatures we start from the AdS$%
\mathcal{L}_{4}$-valued one-form gauge connetion

\begin{equation}
A=\frac{1}{2}\omega ^{ab}J_{ab}+\frac{1}{l}e^{a}P_{a}+\frac{1}{2}k^{ab}Z_{ab}
\label{10}
\end{equation}%
where $a,b=0,1,2,3$ are tangent space indices raised and lowered with the
Minkowski metric $\eta _{ab}$, and where 
\begin{equation*}
e^{a}=e_{\mu }^{a}dx^{\mu },\text{ \ }\omega ^{ab}=\omega _{\mu
}^{ab}dx^{\mu },\text{ \ }k^{ab}=k_{\mu }^{ab}dx^{\mu },
\end{equation*}%
are the $e_{\mu }^{a}$ vierbein, the $\omega _{\mu }^{ab}$ spin connection
and the $k_{\mu }^{ab}$ new non-abelian gauge fields. The generators $%
P_{a},J_{ab},Z_{ab}$ satisfy the AdS$\mathcal{L}_{4}$\textbf{-}algebra
commutation relations (\ref{exp1}).

The corresponding two-form curvature can be obtained from 
\begin{equation}
F=dA+\frac{1}{2}\left[ A,A\right] .  \label{11}
\end{equation}%
In fact, from (\ref{10}) and (\ref{11}) we find 
\begin{equation}
F=\frac{1}{2}R^{ab}J_{ab}+\frac{1}{l}\mathcal{T}^{a}P_{a}+\frac{1}{2}%
F^{ab}Z_{ab},  \label{12}
\end{equation}%
with 
\begin{align}
\mathcal{T}^{a}& =T^{a}+k_{\;c}^{a}e^{c},  \notag \\
R^{ab}& =d\omega ^{ab}+\omega _{\,c}^{a}\omega ^{cb},  \notag \\
F^{ab}& =Dk^{ab}+k_{\;\;c}^{[a}k^{c|b]}+\Lambda e^{a}e^{b},  \label{13}
\end{align}%
where $T^{a}$ and $R^{ab}$ are the standard torsion and curvature, $F^{ab}$
is the curvature of the non-abelian gauge fields $k^{ab}$, $%
Dk^{ab}=dk^{ab}+\omega _{\;e}^{[a}k^{e|b]}$ is the covariant derivative with
respect to $\omega ^{ab}$, and $\Lambda =1/l^{2}$.

The term $k_{\;\;c}^{[a}k^{c|b]}$ comes from the commutator $\left[
Z_{ab},Z_{cd}\right] $, which is zero by Maxwell's algebra. $F^{ab}$
corresponds to the curvature $2$-form associated with gauge fields $k^{ab}$
for the AdS$\mathcal{L}_{4}$ algebra. The term $k_{\;\;b}^{a}e^{b}$ comes
from the commutator $\left[ P_{a},Z_{cd}\right] $, which for the Maxwell
algebra is zero. We can therefore recover the structure equations of
Maxwell's algebra by setting these commutators to zero.

The corresponding Bianchi identities are obtained in the usual form. The
covariant derivative of the generic curvature $F$ is given by 
\begin{equation}
DF=dF+\left[ A,F\right] .  \label{14}
\end{equation}%
From (\ref{10}), (\ref{11}) and (\ref{14}) we find,

\begin{align}
D\mathcal{T}^{\text{ }a}+k_{\;b}^{a}\mathcal{T}^{b}& =\left(
R_{\;b}^{a}+F_{\;b}^{a}\right) e^{b}, \\
DR^{ab}& =0, \\
DF^{ab}+k_{\text{ \ \ }c}^{[a}F^{c|b]}+\Lambda e^{[a}\mathcal{T}^{\text{ }%
b]}+R_{\text{ \ }c}^{[a}k^{a|b]}& =0,
\end{align}%
where $D\mathcal{T}^{\text{ }a}=d\mathcal{T}^{\text{ }a}+\omega _{\;b}^{a}%
\mathcal{T}^{\text{ }b}$. $\ $

Taking into account that the gauge potencials transform as $\delta
_{\varepsilon }A=D\varepsilon ,$ where $\varepsilon $ is a AdS$\mathcal{L}%
_{4}$ algebra valued parameter given by 
\begin{equation}
\varepsilon (x)=\frac{1}{2}\pi ^{ab}J_{ab} + \frac{1}{l}\rho ^{a}P_{a} +%
\frac{1}{2}\xi ^{ab}Z_{ab},  \label{16}
\end{equation}%
it is direct to find%
\begin{eqnarray}
\delta _{\varepsilon }e^{a} &=&D\rho ^{a}+\pi _{\,d}^{a}e^{d}+k_{\text{ \ }%
c}^{a}\rho ^{c}+\xi _{\text{ \ }c}^{a}e^{c}  \label{16'} \\
\delta _{\varepsilon }\omega ^{ab} &=&D\pi ^{ab}  \label{16''} \\
\delta _{\varepsilon }k^{ab} &=&D\xi ^{ab}+k_{\text{ \ \ }c}^{[a}\pi
^{c|b]}+k_{\text{ \ \ }c}^{[a}\xi ^{c|b]}+\Lambda e^{[a}\rho ^{b]}.
\label{16'''}
\end{eqnarray}

On the another hand, from $\delta _{\varepsilon }F=\left[ F,\varepsilon %
\right] $ we find that the curvatures transform as%
\begin{align}
\delta R^{ab}& =R_{\;\;c}^{[a}\pi ^{c|b]}, \\
\delta \mathcal{T}^{a}& =\pi _{\;c}^{a}\mathcal{T}^{c}+R_{\;c}^{a}\rho
^{c}+F_{\;c}^{a}\rho ^{c}+\xi _{\;c}^{a}\mathcal{T}^{c}, \\
\delta F^{ab}& =F_{\;\;c}^{[a}\pi ^{c|b]}+R_{\;\;c}^{[a}\xi
^{c|b]}+F_{\;\;c}^{[a}\xi ^{c|b]}+\Lambda \mathcal{T}^{[a}\rho ^{b]}.
\end{align}

It might be of interest to note that the terms $F_{\;c}^{a}\rho ^{c}$ y $\xi
_{\;c}^{a}\mathcal{T}^{c}$ have their origin in the commutators $\left[
Z_{ab},P_{c}\right] $ and $\left[ P_{a},Z_{cd}\right] $ respectively, which
are null in the case of Maxwell's algebra. Similarly the terms $\pi
_{\;c}^{a}\mathcal{T}^{c},$ $R_{\;c}^{a}\rho ^{c},$ $F_{\;\;c}^{[a}\xi
^{c|b]},$ $F_{\;\;c}^{[a}\pi ^{c|b]},$ $R_{\;\;c}^{[a}\xi ^{c|b]},$\ $\frac{1%
}{l^{2}}\mathcal{T}^{[a}\rho ^{b]},$\ have their origin in the commutators $%
\left[ P_{a},J_{cd}\right] $, $\left[ J_{ab},P_{c}\right] $, $\left[
Z_{ab},Z_{cd}\right] $, $\left[ Z_{ab},J_{cd}\right] ,$ $\left[ J_{ab},Z_{cd}%
\right] ,\left[ P_{a},P_{b}\right] $, which is null for Maxwell's algebra,
respectively.

\subsection{\textbf{Four-dimensional }\textbf{AdS}$\mathcal{L}_{4}$\textbf{%
-action gravity}}

Following the procedure developed in Refs. \cite{azcarr}, \cite{azcarr1} we
postulate the following lagrangians

\begin{equation}
\mathcal{L}_{1}=\varepsilon _{abcd}R^{ab}F^{cd},\text{ \ \ \ }\mathcal{L}%
_{2}=\frac{1}{2}\varepsilon _{abcd}F^{ab}F^{cd},  \label{L2}
\end{equation}%
from where we find that a gravitational action can be constructed from the
lagrangian 
\begin{equation}
-\frac{1}{2\kappa \Lambda }\mathcal{L}_{1}=\mathcal{L}_{E-H}-\frac{1}{%
2\kappa \Lambda }\varepsilon _{abcd}R^{ab}k_{\;\;c}^{[a}k^{c|b]} +\text{%
boundary terms},
\end{equation}%
where%
\begin{equation*}
\mathcal{L}_{E-H}=-\frac{1}{2\kappa }\varepsilon _{abcd}R^{ab}e^{c}e^{d},
\end{equation*}%
is the Einstein-Hilbert term, and the lagrangian

\begin{align}
\frac{\lambda }{2\kappa \Lambda ^{2}}\mathcal{L}_{2}&=\mathcal{L}_{cosm}+%
\frac{\lambda }{4\kappa \Lambda ^{2}}\varepsilon _{abcd}Dk^{ab}Dk^{cd}+\frac{%
\lambda }{2\kappa \Lambda ^{2}}\varepsilon
_{abcd}Dk^{ab}k_{\;\;e}^{[c}k^{e|d]}  \notag \\
& +\frac{\mu }{2\kappa }\varepsilon _{abcd}Dk^{ab}e^{c}e^{d}+\frac{\lambda }{%
4\kappa \Lambda ^{2}}\varepsilon
_{abcd}k_{\;\;f}^{[a}k^{f|b]}k_{\;\;e}^{[c}k^{e|d]}+\frac{\mu }{2\kappa }%
k_{\;\;e}^{[a}k^{e|b]}e^{c}e^{d},
\end{align}%
where $\Lambda \equiv 1/l^{2}$, $\lambda $ is the cosmological constant, $%
\mu =\lambda /\Lambda $ and 
\begin{equation*}
\mathcal{L}_{cosm}=\frac{\lambda }{4\kappa }\varepsilon
_{abcd}e^{a}e^{b}e^{c}e^{d},
\end{equation*}%
is the standard cosmological term. This means $\mathcal{L}_{2}$ can be
identified, following \cite{azcarr}, as an extension of the cosmological
term for the AdS$\mathcal{L}_{4}$ algebra. So that a Lagrangian for the AdS$%
\mathcal{L}_{4}$ gravity is given by 
\begin{align}
\mathcal{L}_{AdS\mathcal{L}_{4}}& =\frac{\mu }{2\kappa \Lambda }\left( -%
\mathcal{L}_{1}+\mu \mathcal{L}_{2}\right) , \\
\mathcal{L}_{AdS\mathcal{L}_{4}}& =\mathcal{L}_{E-H}+\mathcal{L}_{cosm}+%
\frac{\mu }{2\kappa }\varepsilon _{abcd}Dk^{ab}e^{c}e^{d}  \notag \\
& +\frac{\lambda }{4\kappa \Lambda ^{2}}\varepsilon _{abcd}Dk^{ab}Dk^{cd} +%
\frac{\lambda }{2\kappa \Lambda ^{2}}\varepsilon
_{abcd}Dk^{ab}k_{\;\;e}^{[c}k^{e|d]}-\frac{1}{2\kappa \Lambda }\varepsilon
_{abcd}R^{ab}k_{\;\;e}^{[c}k^{e|d]}  \notag \\
& +\frac{\lambda }{4\kappa \Lambda ^{2}}\varepsilon
_{abcd}k_{\;\;f}^{[a}k^{f|b]}k_{\;\;e}^{[c}k^{e|d]} +\frac{\mu }{2\kappa }%
\varepsilon _{abcd}k_{\;\;e}^{[a}k^{e|b]}e^{c}e^{d},  \label{act-adsl4}
\end{align}%
which, after some algebra, takes the form

\begin{align}
\mathcal{L}& =\mathcal{L}_{E-H}+\mathcal{L}_{cosm}+\frac{\mu }{2\kappa}%
\varepsilon _{abcd}Dk^{ab}e^{c}e^{d}+\frac{\lambda }{4\kappa\Lambda ^{2}}%
\varepsilon _{abcd}Dk^{ab}Dk^{cd}  \notag \\
& -\frac{\mu }{\kappa \Lambda }\varepsilon _{abcd}R^{ab}k_{\;\;e}^{c}k^{ed}+%
\frac{\lambda }{\kappa\Lambda ^{2}}\varepsilon
_{abcd}k_{\;f}^{a}k^{fb}k_{\;e}^{c}k^{ed}+\frac{\mu }{\kappa}\varepsilon
_{abcd}k_{\;e}^{a}k^{eb}e^{c}e^{d}  \notag \\
& +\frac{\lambda }{\kappa \Lambda ^{2}}\varepsilon
_{abcd}Dk^{ab}k_{\;\;e}^{c}k^{ed}.  \label{g4}
\end{align}

Here, we can see that the four first terms of (\ref{g4}) correspond to the
lagrangian for the Maxwell algebra (see equation (\ref{max}) below).

\subsection{\textbf{Field Equations for }\textbf{AdS}$\mathcal{L}_{4}$-%
\textbf{extended Einstein gravity}}

The variation of the action \eqref{g4} with respect to the vielbein $e^{a}$
leads to

\begin{equation}
\delta _{e^{a}}\mathcal{L}=\frac{1}{\kappa }\varepsilon _{abcd}\left( \mu %
\left[ Dk^{ab}+\Lambda e^{a}e^{b}+k_{\;\;e}^{[a}k^{e|b]}\right]
e^{c}-R^{ab}e^{c}\right) \delta e^{d}=0,
\end{equation}%
i.e., leads to the field equation 
\begin{equation}
\varepsilon _{abcd}\left( \mu F^{ab}e^{c}-R^{ab}e^{c}\right) =0,  \label{eq1}
\end{equation}%
where $F^{ab}$ is given by the last equation of \ (\ref{13}).

The variation of the action \eqref{g4} with respect to the spin connection $%
\omega ^{ab}$ leads to

\begin{equation}
\delta _{\omega ^{ab}}\mathcal{L}=\delta \omega ^{ab}\left\{ -\frac{1}{%
\kappa }\varepsilon _{abcd}\left[ De^{c}e^{d}-\frac{\mu }{\Lambda }%
k_{\;\;e}^{c}\left( Dk^{ed}+k_{\;\;f}^{[e}k^{f|d]}+\Lambda e^{e}e^{d}\right) %
\right] \right\} =0  \notag
\end{equation}%
so that, the field equation corresponding to the variation with respect to $%
\omega ^{ab}$ is

\begin{equation}
\frac{1}{\kappa }\varepsilon _{abcd}\left( T^{c}e^{d}-\frac{\mu }{\Lambda }%
k_{\;\;e}^{c}F^{ed}\right) =0  \label{eq2}
\end{equation}

The variation of the action \eqref{g4} with respect to the vielbein $k^{ab}$
leads to

\begin{align}
\delta _{k^{ab}}\mathcal{L}& =-\frac{\mu }{\kappa }\varepsilon _{abcd}\delta
k^{ab}De^{c}e^{d}-\frac{\mu }{2\kappa \Lambda }\varepsilon
_{abcd}DDk^{ab}\delta k^{cd}  \notag \\
& -\frac{2\mu }{\kappa \Lambda }\varepsilon
_{abcd}R^{ab}k_{\,\,\,e}^{c}\delta k^{ed}+\frac{2\mu }{\kappa \Lambda }%
\varepsilon _{abcd}k_{\,\,\,f}^{\left[ a\right. }k^{\left. f|b\right]
}k_{\,\,\,e}^{c}\delta k^{ed}  \notag \\
& +\frac{2\mu }{\kappa }\varepsilon _{abcd}k_{\,\,\,e}^{a}\delta
k^{eb}e^{c}e^{d}+\frac{2\mu }{\kappa \Lambda }\varepsilon
_{abcd}Dk^{ab}k_{\,\,\,e}^{c}\delta k^{ed}  \notag \\
& +\text{boundary terms}=0,
\end{align}%
and therefore the corresponding field equation is given by

\begin{equation}
\frac{1}{\kappa }\varepsilon _{abcd}\left( T^{c}e^{d}-\frac{3}{2}\frac{\mu }{%
\Lambda }R^{ce}k_{\,\,\,e}^{d}-\frac{2\mu }{\Lambda }k_{\,\,\,e}^{c}F^{ed}%
\right) =0,  \label{eq3}
\end{equation}%
where $T^{c}$, $R^{ce}$ and $F^{ed}$ are given by (\ref{13}).

\section{\textbf{AdS}$\mathcal{L}_{5}$\textbf{-generalizacion of the
Einstein--Hilbert-Cartan gravity}}

Now we will study the generalizacion of Einstein gravity considering AdS$%
\mathcal{L}_{5}$ symmetries.

\subsection{\textbf{Geometric aspect of the gauging of the }\textbf{AdS}$%
\mathcal{L}_{5}$\textbf{-algebra}}

In this case, the AdS$\mathcal{L}_{5}$-valued one-form gauge connetion is
given by

\begin{equation*}
A=\frac{1}{2}\omega ^{ab}J_{ab}+\frac{1}{l}e^{a}P_{a}+\frac{1}{2}%
k^{ab}Z_{ab}+\frac{1}{l}h^{a}Z_{a},
\end{equation*}%
where now the generators $P_{a},J_{ab},Z_{ab},Z_{a}$ satisfy the AdS$%
\mathcal{L}_{5}$\textbf{-}algebra commutation relations (\ref{exp2}).

Following the same procedure of the previous section we find that the
corresponding curvature $2$-form is given by

\begin{equation}
F=\frac{1}{2}\mathcal{R}^{ab}J_{ab}+\frac{1}{l}\mathcal{\tilde{T}}^{a}P_{a}+%
\frac{1}{2}\tilde{F}^{ab}Z_{ab}+\frac{1}{l}H^{a}Z_{a}  \notag
\end{equation}%
with 
\begin{eqnarray}
\mathcal{R}^{ab} &=&R^{ab}+k_{\,c}^{a}k^{cb}+2\Lambda e^{a}h^{b}  \label{26'}
\\
\mathcal{\tilde{T}}^{a} &=&T^{a}+k_{\,b}^{a}h^{b}=De^{a}+k_{\,b}^{a}h^{b}
\label{26''} \\
\tilde{F}^{ab} &=&Dk^{ab}+\Lambda h^{a}h^{b}+\Lambda e^{a}e^{b}
\label{26'''} \\
H^{a} &=&Dh^{a}+k_{\,b}^{a}e^{b}.  \label{26''''}
\end{eqnarray}%
where $T^{a}$ and $R^{ab}$ are the standard torsion and curvature, $\Tilde{F}%
^{ab}$ is the curvature of the non-abelian gauge fields $k^{ab}$ with $%
Dk^{ab}=dk^{ab}+\omega _{\;e}^{[a}k^{e|b]}$ and $H^{a}$ is the curvature
associated to the generator $Z_{a}$.

The covariant derivative of the curvatures lead to the following Bianchi
identities

\begin{equation}
D\mathcal{R}^{ab}+k_{\;\;c}^{[a}\tilde{F}^{c|b]}+\Lambda h^{[a}\mathcal{%
\tilde{T}}^{b]}+\Lambda e^{[a}H^{b]}=0,
\end{equation}

\begin{equation}
D\mathcal{\tilde{T}}^{a}+\mathcal{R}_{\;\;b}^{a}e^{b}+\tilde{F}%
_{\;\;b}^{a}h^{b}+k_{\;\;b}^{a}H^{b}=0,
\end{equation}

\begin{equation}
D\tilde{F}^{ab}+\mathcal{R}_{\;\;c}^{[a}k^{c|b]}+\Lambda e^{[a}\mathcal{%
\tilde{T}}^{b]}+\Lambda h^{[a}H^{b]}=0,
\end{equation}

\begin{equation}
DH^{a}+\mathcal{R}_{\;\;b}^{a}h^{b}+k_{\;b}^{a}\mathcal{\tilde{T}}^{b}+%
\tilde{F}_{\;b}^{a}e^{b}=0,
\end{equation}%
where $D\mathcal{T}^{a}=d\mathcal{T}^{a}+\omega _{\;b}^{a}\mathcal{T}^{b}$.

The gauge potencials transform as 
\begin{equation}
\delta _{\varepsilon }A=D\varepsilon =d\varepsilon +\left[ A,\varepsilon %
\right] ,
\end{equation}%
where $\varepsilon $ is a AdS$\mathcal{L}_{5}$ algebra valued parameter
given by 
\begin{equation}
\varepsilon =\frac{1}{2}\pi ^{ab}J_{ab}+\frac{1}{l}\rho ^{a}P_{a}+\frac{1}{2}%
\xi ^{ab}Z_{ab}+\frac{1}{l}\sigma ^{a}Z_{a}.
\end{equation}%
Using the commutation relation of the AdS$\mathcal{L}_{5}$\textbf{-}algebra
we find

\begin{align}
\delta \omega ^{ab}& =d\pi ^{ab}+\omega _{\;\;c}^{[a}\pi
^{c|b]}+k_{\;\;c}^{[a}\xi ^{c|b]}+\Lambda e^{[a}\sigma ^{b]}+\Lambda
h^{[a}\rho ^{b]}, \\
\delta e^{a}& =d\rho ^{a}+\pi _{\;c}^{a}e^{c}+\omega _{\;c}^{a}\rho ^{c}+\xi
_{\;c}^{a}h^{c}+k_{\;c}^{a}\sigma ^{c}, \\
\delta k^{ab}& =d\xi ^{ab}+k_{\;\;c}^{[a}\pi ^{c|b]}+\omega _{\;\;c}^{[a}\xi
^{c|b]}+\Lambda e^{[a}\rho ^{b]}+\Lambda h^{[a}\sigma ^{b]}, \\
\delta h^{a}& =d\sigma ^{a}+\pi _{\;c}^{a}h^{c} +k_{\;c}^{a}\rho ^{c}+\xi
_{\;c}^{a}e^{c}+\omega _{\;c}^{a}\sigma ^{c}.
\end{align}

On the another hand, from 
\begin{equation*}
\delta _{\varepsilon }F=\left[ F,\varepsilon \right]
\end{equation*}%
we find that the curvatures transform as

\begin{align}
\delta \mathcal{R}^{ab}& =\mathcal{R}_{\;\;c}^{[a}\pi ^{c|b]}+\tilde{F}%
_{\;\;c}^{[a}\xi ^{c|b]}+\Lambda H^{[a}\rho ^{b]}+\Lambda\mathcal{\tilde{T}}%
^{[a}\sigma ^{b]}, \\
\delta \mathcal{\tilde{T}}^{a}& =\pi _{\;c}^{a}\mathcal{\tilde{T}}^{c}+%
\mathcal{R}_{\;c}^{a}\rho ^{c}+\xi _{\;c}^{a}H^{c}+\tilde{F}_{\;c}^{a}\sigma
^{c}, \\
\delta \tilde{F}^{ab}& =\tilde{F}_{\;\;c}^{[a}\pi ^{c|b]}+\mathcal{R}%
_{\;\;c}^{[a}\xi ^{c|b]}+\Lambda \mathcal{\tilde{T}}^{[a}\rho ^{b]}+\Lambda
H^{[a}\sigma ^{b]}, \\
\delta H^{a}& =\pi _{\;c}^{a}H^{c}+\tilde{F}_{\;c}^{a}\rho ^{c}+\xi
_{\;c}^{a}\mathcal{\tilde{T}}^{c}+\mathcal{R}_{\;c}^{a}\sigma ^{c}.
\end{align}

\subsection{\textbf{Four-dimensional }\textbf{AdS}$\mathcal{L}_{5}$\textbf{%
-Einstein-Hilbert action}}

The same procedure of the previous section leads to an action for the AdS$%
\mathcal{L}_{5}$-gravity constructed from the following 4-forms invariant%
\begin{equation}
\mathcal{L}_{1}=\varepsilon _{abcd}\mathcal{R}^{ab}\wedge \tilde{F}^{cd},%
\hspace{1.5cm}\mathcal{L}_{2}=\frac{1}{2}\varepsilon _{abcd}\tilde{F}%
^{ab}\wedge \tilde{F}^{cd}.
\end{equation}

\begin{eqnarray}
-\frac{1}{2\kappa \Lambda }\mathcal{L}_{1} &=&-\frac{1}{2\kappa \Lambda }%
\varepsilon _{abcd}\mathcal{R}^{ab}\wedge \tilde{F}^{cd}  \notag \\
&=&\mathcal{L}_{E-H}-\frac{1}{2\kappa}\varepsilon_{abcd}R^{ab}h^{c}h^{d}+%
\frac{\mu}{2\kappa\lambda} \varepsilon_{abcd}k^{a}_{\,\,\,e}k^{eb}Dk^{cd} 
\notag \\
&&-\frac{1}{2\kappa}\varepsilon_{abcd}k^{a}_{\,\,\,e}k^{eb}e^{c}e^{d} - 
\frac{1}{2\kappa}\varepsilon_{abcd}k^{a}_{\,\,\,e}k^{eb}h^{c}h^{d}-\frac{1}{%
\kappa}\varepsilon_{abcd}e^{a}h^{b}Dk^{cd}  \notag \\
&&-\frac{\Lambda}{\kappa}\varepsilon_{abcd}e^{a}h^{b}e^{c}e^{d} -\frac{%
\Lambda}{\kappa}\varepsilon_{abcd}e^{a}h^{b}h^{c}h^{d},
\end{eqnarray}

\begin{eqnarray}
\frac{\lambda }{2\kappa \Lambda ^{2}}\mathcal{L}_{2} &=&\frac{\lambda }{%
4\kappa \Lambda ^{2}}\varepsilon _{abcd}\tilde{F}^{ab}\wedge \tilde{F}^{cd} 
\notag \\
&=&\mathcal{L}_{cosm}+\frac{\lambda }{4\kappa \Lambda ^{2}}\varepsilon
_{abcd}Dk ^{ab}Dk ^{cd} +\frac{\lambda }{2\kappa \Lambda }\varepsilon
_{abcd}Dk ^{ab}h^{c}h^{d}  \notag \\
&+&\frac{\lambda }{2\kappa \Lambda }\varepsilon _{abcd}Dk ^{ab}e^{c}e^{d}+%
\frac{\lambda }{2\kappa }\varepsilon _{abcd}e^{a}e^{b}h^{c}h^{d}+ \frac{%
\lambda }{4\kappa }\varepsilon _{abcd}h^{a}h^{b}h^{c}h^{d},
\end{eqnarray}

Defining $\mu =\lambda /\Lambda $ we propose 
\begin{align}
\mathcal{L}_{\text{AdS}\mathcal{L}_{5}}& =\mathcal{L}_{E-H}+\mathcal{L}%
_{cosm}-\frac{1}{2\kappa }\varepsilon _{abcd}R^{ab}h^{c}h^{d}-\frac{\mu }{%
2\kappa \lambda }\varepsilon _{abcd}k_{\,\,\,e}^{a}k^{eb}Dk^{cd}  \notag \\
& -\frac{1}{2\kappa }\varepsilon _{abcd}k_{\,\,\,e}^{a}k^{eb}e^{c}e^{d}-%
\frac{1}{2\kappa }\varepsilon _{abcd}k_{\,\,\,e}^{a}k^{eb}h^{c}h^{d}-\frac{1%
}{\kappa }\varepsilon _{abcd}e^{a}h^{b}Dk^{cd}  \notag \\
& -\frac{\Lambda }{\kappa }\varepsilon _{abcd}h^{a}e^{b}e^{c}e^{d}-\frac{%
\Lambda }{\kappa }\varepsilon _{abcd}e^{a}h^{b}h^{c}h^{d}+\frac{\mu ^{2}}{%
4\kappa \lambda }\varepsilon _{abcd}Dk^{ab}Dk^{cd}  \notag \\
& +\frac{\lambda }{2\kappa \Lambda }\varepsilon _{abcd}Dk^{ab}e^{c}e^{d}+%
\frac{\lambda }{2\kappa \Lambda }\varepsilon _{abcd}Dk^{ab}h^{c}h^{d}  \notag
\\
& +\frac{\lambda }{2\kappa }\varepsilon _{abcd}e^{a}e^{b}h^{c}h^{d}+\frac{%
\lambda }{4\kappa }\varepsilon _{abcd}h^{a}h^{b}h^{c}h^{d},  \label{g4'}
\end{align}

where,

\begin{align}
\mathcal{L}_{E-H}& =-\frac{1}{2\kappa }\varepsilon _{abcd}R^{ab}e^{c}e^{d},
\\
\mathcal{L}_{cosm}& =\frac{\lambda }{4\kappa }\varepsilon
_{abcd}e^{a}e^{b}e^{c}e^{d}
\end{align}

\subsection{\textbf{Field Equations for }\textbf{AdS}$\mathcal{L}_{5}$-%
\textbf{extended Einstein gravity}}

The variation of $\mathcal{L}_{\text{AdS}\mathcal{L}_{5}}$ with respect to $%
e^{a}$ leads to the following equations of motion

\begin{align}
\delta _{e}\mathcal{L}_{\text{AdS}\mathcal{L}_{5}}& =\left[ L_{\text{AdS}%
\mathcal{L}_{5}}\right] _{e^{a}}\delta e^{d}  \notag \\
& =\frac{1}{\kappa }\varepsilon _{abcd}\left\{
-\left(R^{ab}+k_{\;\;c}^{a}k^{cb}+2\Lambda e^{a}h^{b}\right)e^{c}-\left(
Dk^{ab}+\Lambda e^{a}e^{b}+\Lambda h^{a}h^{b}\right) h^{c}\right.  \notag \\
& \left. +\mu \left( Dk^{ab}+\Lambda h^{a}h^{b}+\Lambda e^{a}e^{b}\right)
e^{c}\right\} \delta e^{d},
\end{align}

so that

\begin{equation}
\left[ L_{\text{AdS}\mathcal{L}_{5}}\right] _{e^{a}}=\frac{1}{\kappa }%
\varepsilon _{abcd}\left\{ -\mathcal{R}^{ab}e^{c}-\Tilde{F}^{ab}h^{c}+\mu 
\Tilde{F}^{ab}e^{c}\right\} =0  \label{eq4}
\end{equation}

The variation of $\mathcal{L}_{\text{AdS}\mathcal{L}_{5}}$ with respect to $%
h^{a}$ leads to the following equations of motion

\begin{align}
\delta _{h}\mathcal{L}_{\text{AdS}\mathcal{L}_{5}}& =\left[ L_{\text{AdS}%
\mathcal{L}_{5}}\right] _{h^{a}}\delta h^{d}  \notag \\
& =\frac{1}{\kappa }\varepsilon _{abcd}\left\{ -\left(
R^{ab}+k_{\;\;e}^{a}k^{eb}+2\Lambda e^{a}h^{b}\right) h^{c}-\left(
Dk^{ab}+\Lambda e^{a}e^{b}+\Lambda h^{a}h^{b}\right) e^{c}\right.  \notag \\
& \left. +\mu \left( Dk^{ab}+\Lambda e^{a}e^{b}+\Lambda h^{a}h^{b}\right)
h^{c}\right\} \delta h^{d},
\end{align}

i.e.,%
\begin{equation}
\left[ L_{\text{AdS}\mathcal{L}_{5}}\right] _{h^{a}}=\frac{1}{\kappa }%
\varepsilon _{abcd}\left( -\mathcal{R}^{ab}h^{c}-\tilde{F}^{ab}e^{c}+\mu 
\tilde{F}^{ab}h^{c}\right) .  \label{eq5}
\end{equation}%
The variation of $\mathcal{L}_{\text{AdS}\mathcal{L}_{5}}$ with respect to $%
\omega ^{ab}$ leads to the following equations of motion

\begin{align}
\delta _{\omega }\mathcal{L}_{\text{AdS}_{\mathcal{L}_{5}}}& =\delta \omega
^{ab}\left[ L_{\text{AdS}\mathcal{L}_{5}}\right] _{\omega ^{ab}}  \notag \\
& =\frac{1}{\kappa }\varepsilon _{abcd}\delta \omega ^{ab}\left\{
-T^{c}e^{d}-Dh^{c}h^{d}-\frac{\mu ^{2}}{\lambda }k_{\;\;e}^{c}\tilde{F}^{ed}-%
\frac{\mu }{\lambda }k_{\;\;e}^{c}\left( k_{\;\;f}^{e}k^{fd}+2\Lambda
e^{e}h^{d}\right) \right\} .
\end{align}

From (\ref{26'}) we know that $\mathcal{R}^{ed}-R^{ed}=k_{\,c}^{e}k^{cd}+2%
\Lambda e^{e}h^{d} $, so 
\begin{equation*}
\varepsilon _{abcd}k_{\;\;e}^{c}\left( \mathcal{R}^{ed}-R^{ed}\right)
=\varepsilon _{abcd}k_{\;\;e}^{c}\left[ k_{\,\text{\ }f}^{e}k^{fd}+2\Lambda
e^{e}h^{d}\right] ,
\end{equation*}%
and therefore%
\begin{eqnarray}
\left[ L_{\text{AdS}\mathcal{L}_{5}}\right] _{\omega ^{ab}} &=&\frac{1}{%
\kappa }\varepsilon _{abcd}\delta \omega ^{ab}\left( -T^{c}e^{d}-Dh^{c}h^{d}-%
\frac{\mu }{\lambda }k_{\;\;e}^{c}\mathcal{R}^{ed}+\frac{\mu }{\lambda }%
k_{\;\;e}^{c}R^{ed}-\frac{\mu ^{2}}{\lambda }k_{\;\;e}^{c}\Tilde{F}%
^{ed}\right) =0  \notag \\
&&  \label{eq6}
\end{eqnarray}

The variation of $\mathcal{L}_{\text{AdS}\mathcal{L}_{5}}$ with respect to $%
k^{ab}$ leads to field equation

\begin{align}
\delta _{k}\mathcal{L}_{\text{AdS}\mathcal{L}_{5}}& =\delta k^{ab}\left[ L_{%
\text{AdS}\mathcal{L}_{5}}\right] _{\delta k^{ab}}  \notag \\
& =\frac{1}{\kappa }\varepsilon _{abcd}\delta k^{ab}\left( \frac{2\mu }{%
\lambda }k_{\;\;e}^{c}Dk^{ed}+k_{\;\;e}^{c}e^{e}e^{d}+k_{\;%
\;e}^{c}h^{e}h^{d}-Dh^{c}e^{d}\right.  \notag \\
& \left. -De^{c}h^{d}+\mu De^{c}e^{d}+\mu Dh^{c}h^{d}-\frac{\mu ^{2}}{%
2\lambda }R_{\;\;e}^{c}k^{ed}\right) ,
\end{align}

so that,

\begin{align}
& \left[ L_{\text{AdS}\mathcal{L}_{5}}\right] _{\delta k^{ab}}  \notag \\
& =\frac{1}{\kappa }\varepsilon _{abcd}\left( \frac{\mu }{\lambda }%
k_{\;\;e}^{c}\tilde{F}^{ed}+\frac{\mu }{\lambda }k_{\;\;e}^{c}Dk^{ed}-\frac{%
\mu ^{2}}{2\lambda }k_{\;\;e}^{c}R^{ed}\right.  \notag \\
& \left. +Dh^{c}\left( \mu h^{d}-e^{d}\right) +De^{c}\left( \mu
e^{d}-h^{d}\right) \right) =0  \label{eq7}
\end{align}

\section{$\mathfrak{B}_{4}$ \textbf{and} $\mathfrak{B}_{5}$ \textbf{actions
from }$\mathbf{AdS}\mathcal{L}_{4}$ \textbf{and} $\mathbf{AdS}\mathcal{L}%
_{5} $ \textbf{gravities}}

In this Section we obtain the well-known Maxwell gravity from the AdS$%
\mathcal{L}_{4}$ action and a four-dimensional gravity action from the AdS$%
\mathcal{L}_{5}$ gravity action together with their corresponding field
equations.

\subsection{\textbf{Maxwell gravity from }\textbf{AdS}$\mathcal{L}_{4}$%
\textbf{-action}}

From (\ref{g4}) we see that the Lagrangian AdS$\mathcal{L}_{4}$ differs from
the Maxwell Lagrangian, of Refs. \cite{azcarr}, \cite{azcarr1}, in the
following four terms

\begin{equation}
+\frac{\lambda }{\kappa \Lambda ^{2}}\varepsilon
_{abcd}Dk^{ab}k_{\;\;e}^{c}k^{ed}-\frac{\mu }{\kappa\Lambda }\varepsilon
_{abcd}R^{ab}k_{\;\;e}^{c}k^{ed}+\frac{\lambda }{\kappa\Lambda ^{2}}%
\varepsilon _{abcd}k_{\;f}^{a}k^{fb}k_{\;e}^{c}k^{ed}+\frac{\mu }{\kappa}%
\varepsilon _{abcd}k_{\;e}^{a}k^{eb}e^{c}e^{d},
\end{equation}%
coming from the commutators $\left[ Z_{ab},P_{c}\right] $ and $\left[
Z_{ab},Z_{cd}\right] $, which in the case of Maxwell's algebra commute. The
natural question is how to obtain the Lagrangian corresponding to the
Maxwell algebra from the Lagrangian for the AdS$\mathcal{L}_{4}$ algebra?
Just as it is possible to obtain the Maxwell algebra from the AdS$\mathcal{L}%
_{4}$ algebra by means of an Inonu-Wigner contraction in the Weimar-Woods
sense, it is also possible to find the Lagrangian for the Maxwell algebra.
Indeed, carrying out the rescaling of the generators $P_{a}\rightarrow \xi
P_{a}$, $Z_{ab}\rightarrow \xi ^{2}Z_{ab}$ and of the fields $%
e^{a}\rightarrow \xi ^{-1}e^{a}$, $k^{ab}\rightarrow \xi ^{-2}k^{ab}$ in the
Lagrangian (\ref{act-adsl4}) for the AdS$\mathcal{L}_{4}$ algebra, we obtain

\begin{equation}
\mathcal{L}=\mathcal{L}_{E-H}+\mathcal{L}_{cosm}+\frac{\mu }{2\kappa }%
\varepsilon _{abcd}Dk^{ab}e^{c}e^{d}+\frac{\mu ^{2}}{4\kappa \lambda }%
\varepsilon _{abcd}D_{\omega }k^{ab}D_{\omega }k^{cd},  \label{max}
\end{equation}%
result that coincides with equation ($29$) of the reference \cite{azcarr}.
It is straightforward to see that the application of the In\"{o}n\"{u}%
-Wigner contraction procedure to the AdS$\mathcal{L}_{4}$-field equations
leads to the equations ($31$), ($34$) and ($37$) of the reference \cite%
{azcarr}.

\subsection{$\mathfrak{B}_{5}$\textbf{\ gravity from }\textbf{AdS}$\mathcal{L%
}_{5}$\textbf{-gravity}}

Considering that the generalized Poincare algebra $\mathfrak{B}_{5}$ can be
obtained from the AdS$\mathcal{L}_{5}$ algebra, by means of an Inonu-Wigner
contraction in the Weimar-Woods sense, the natural question is how to obtain
the corresponding Lagrangian to the $\mathfrak{B}_{5}$ algebra from the
Lagrangian for the AdS$\mathcal{L}_{5}$ algebra?. Following the same
procedure of the previous section we carry out the rescaling of the
generators $P_{a}\rightarrow \xi P_{a}$, $Z_{ab}\rightarrow \xi ^{2}Z_{ab},$ 
$Z_{a}\rightarrow \xi ^{3}Z_{a}$ and of the fields $e^{a}\rightarrow \xi
^{-1}e^{a}$, $k^{ab}\rightarrow \xi ^{-2}k^{ab},$ $h^{a}\rightarrow \xi
^{-3}h^{a}$ in the Lagrangian (\ref{g4'}) for the AdS$\mathcal{L}_{5}$
algebra, we obtain the four-dimensional Lagrangian for $\mathfrak{B}_{5}$
algebra%
\begin{eqnarray}
\mathcal{L}_{\mathfrak{B}_{5}}^{(4D)} &=&\mathcal{L}_{E-H}+\mathcal{L}%
_{cosm}+\frac{\lambda }{2\kappa \Lambda }\varepsilon
_{abcd}Dk^{ab}e^{c}e^{d}+\frac{\mu ^{2}}{4\kappa \lambda }\varepsilon
_{abcd}Dk^{ab}Dk^{cd}  \notag \\
&&-\frac{1}{2\kappa }\varepsilon _{abcd}R^{ab}h^{c}h^{d}-\frac{1}{\kappa}%
\varepsilon _{abcd}Dk^{ab}h^{c}e^{d}-\frac{\Lambda }{\kappa }\varepsilon
_{abcd}h^{a}e^{b}e^{c}e^{d}  \notag \\
&&-\frac{\Lambda }{\kappa }\varepsilon _{abcd}e^{a}h^{b}h^{c}h^{d}+\frac{%
\lambda }{2\kappa \Lambda }\varepsilon _{abcd}Dk^{ab}h^{c}h^{d}+\frac{%
\lambda }{2\kappa }\varepsilon _{abcd}e^{a}e^{b}h^{c}h^{d}  \notag \\
&&+\frac{\lambda }{4\kappa }\varepsilon _{abcd}h^{a}h^{b}h^{c}h^{d},
\end{eqnarray}

which contains the Lagrangian corresponding to the $\mathfrak{B}_{5}$
algebra.

Applying the In\"{o}n\"{u}-Wigner contraction procedure to the AdS$\mathcal{L%
}_{5}$-field equations we obtain

\begin{equation*}
\frac{1}{\kappa }\varepsilon _{abcd}\left\{- R^{ab}e^{c} - 2\Lambda
e^{a}h^{b}e^{c} - \Tilde{F}^{ab}h^{c} + \mu \Tilde{F}^{ab}e^{c}\right\} =0
\end{equation*}

\begin{equation*}
\frac{1}{\kappa }\varepsilon _{abcd}\left\{ -R^{ab}h^{c} - 2\Lambda
e^{a}h^{b}h^{c} - \Tilde{F}^{ab}e^{c} + \mu \Tilde{F}^{ab}h^{c}\right\} =0
\end{equation*}

\begin{equation*}
\frac{1}{\kappa }\varepsilon _{abcd}\left\{ -T^{c}e^{d}-D_{\omega}h^{c}h^{d}-%
\frac{\mu ^{2}}{\lambda }k_{\;\;e}^{c}\Tilde{F}^{ed}-2\Lambda
^{2}k_{\;\;e}^{c}e^{e}h^{d}\right\} =0
\end{equation*}

\begin{equation}
\frac{1}{\kappa }\varepsilon _{abcd}\left( Dh^{c}\left( \mu
h^{d}-e^{d}\right) +De^{c}\left( \mu e^{d}-h^{d}\right) -\frac{\mu ^{2}}{%
2\lambda }R_{\;\;e}^{c}k^{ed}\right) =0,
\end{equation}%
which correspond to the field equations for $\mathfrak{B}_{5}$-extended
Einstein gravity.

\section{\textbf{Concluding Remarks}}

In this article we have considered local gauge theories based on the $AdS%
\mathcal{L}_{4}$ and $AdS\mathcal{L}_{5}$ algebras with vierbein, spin
conection and six addicional geometric $k_{\mu }^{ab}$ non-Abelian gauge
fields in the first case and with ten addicional geometric ($k_{\mu
}^{ab},h_{\mu }^{a}$) non-Abelian gauge fields, in the second case.

The geometry of space-times based on the above mentioned algebras involve
new curvature tensors that allow to construct new gravity actions which lead
to modifications of the Einstein gravity.

We have constructed the curvatures $2$-form asociated with $AdS\mathcal{L}%
_{4}$ and $AdS\mathcal{L}_{5}$ valued one-form gauge connections, which
allow us to construct four-dimensional gravities that generalize the
Einstein Hilbert gravity. From these gravitational actions we find that the
Maxwell extension as well as the $\mathfrak{B}_{5}$ extension of Einstein
gravity can be obtained using the In\"{o}n\"{u}-Wigner contraction method.

The field equations (\ref{eq1}) and (\ref{eq2}) allow us to express the spin
conection as a function of the vierbein $e^{a}$ and the new non-Abelian
gauge fields $k_{\mu }^{ab}.$ This would allow us to obtain a second order
formulation for the $AdS\mathcal{L}_{4}$-gravity, with independent fields $%
e^{a}$ and $k_{\mu }^{ab}$. Similarly, from the equations (\ref{eq4}), (\ref%
{eq5}) and (\ref{eq6}) it might be possible to express the spin connection
as a function of the fields $e^{a}$, $k_{\mu }^{ab}$ and $h_{\mu }^{a}$ and
obtain a second order formulation for the $AdS\mathcal{L}_{5}$-gravity, with
independent fields $e^{a}$, $k_{\mu }^{ab}$ and $h_{\mu }^{a}$.

It might be of interest to note that some years ago, another generalization
of the Einstein-Hilbert-Cartan action, invariant under $AdSL_{4}$, i.e.,
invariant under the local non-Abelian gauge symmetries associated with the $%
Z_{ab}$\ generators, was proposed in Ref. \cite{durka}. The Lagrangian $%
\left( 33\right) $ of this reference coincides with the Lagrangian (\ref{g4}%
) only in the Einstein-Hilbert and the cosmological terms, but it does not
contain terms involving the field $k_{\mu }^{ab}$\ ($h_{\mu }^{ab}$\ in Ref. 
\cite{durka}), i.e. it contains no terms that could contribute to a
cosmological term. The reason for this, according to \cite{durka}, would be
that the starting point for the construction of action $\left( 33\right) $\
was the\ so called BF theory, which is a geometric theory.

In this context, it might also be of interest to note that the
Randall-Sundrum compactification procedure could allow obtaining the action $%
\left( 22\right) $\ from the five-dimensional Chern-Simons gravity invariant
under $AdSL_{4}$.\textbf{\ }This procedure could also be used to obtain the
Lagrangian $\left( 29\right) $\ of reference \cite{azcarr} from the Maxwell
Chern-Simons gravity action \cite{dmos}.

\textbf{Acknowledgments:} This work was supported in part by\textit{\ }%
FONDECYT Grants\textit{\ }No.\textit{\ }1180681 (J.D and P.S) and No.\textit{%
\ }1211219 (P.S) from the Government of Chile. Two of the authors (L.C and
D.S) were supported by Universidad de Concepci\'{o}n, Chile. \ 

\section{\textbf{Appendix A: Generalized AdS and Poincare algebras}}

\subsection{\textbf{AdS}$\mathcal{L}_{4}$, \textbf{AdS}$\mathcal{L}_{5}$ 
\textbf{and} \textbf{AdS}$\mathcal{L}_{6}$\textbf{-algebra}}

\textbf{AdS}$\mathcal{L}_{4}$\textbf{-algebra: \ }The $S$-expansion of the
Lie AdS algebra using as semigroup $S_{\mathcal{M}}^{\left( 2\right)
}=\left\{ \lambda _{0},\lambda _{1},\lambda _{2}\right\} $ endowed with the
multiplication rule \ $\lambda _{\alpha }\lambda _{\beta }=\lambda _{\alpha
+\beta }$ if \ $\alpha +\beta \leq 2;$ $\lambda _{\alpha }\lambda _{\beta }=$
$\lambda _{\alpha +\beta -2}$ if \ $\alpha +\beta >2$, lead us to the so
called AdS$\mathcal{L}_{4}$-algebra, whose generators satisfy the following
commutation relations%
\begin{align}
\left[ J_{ab},J_{cd}\right] & =\eta _{bc}J_{ad}+\eta _{ad}J_{bc}-\eta
_{ac}J_{bd}-\eta _{bd}J_{ac},  \notag \\
\left[ J_{ab},Z_{cd}\right] & =\eta _{bc}Z_{ad}+\eta _{ad}Z_{bc}-\eta
_{ac}Z_{bd}-\eta _{bd}Z_{ac},  \notag \\
\left[ Z_{ab},Z_{cd}\right] & =\eta _{bc}Z_{ad}+\eta _{ad}Z_{bc}-\eta
_{ac}Z_{bd}-\eta _{bd}Z_{ac},  \notag \\
\left[ J_{ab},P_{c}\right] & =\eta _{bc}P_{a}-\eta _{ac}P_{b},\text{ \ \ \ }%
\left[ P_{a},P_{b}\right] =Z_{ab},  \notag \\
\left[ Z_{ab},P_{c}\right] & =\eta _{bc}P_{a}-\eta _{ac}P_{b}.  \label{exp1}
\end{align}%
This algebra was also reobtained in Ref. \cite{gomis} from Maxwell algebra
through a procedure known as deformation.

\textbf{AdS}$\mathcal{L}_{5}$\textbf{-algebra: }the same procedure, but now
using the semigroup $S_{\mathcal{M}}^{\left( 3\right) }=%
\mathbb{Z}
_{4}=\left\{ \lambda _{0},\lambda _{1},\lambda _{2},\lambda _{3}\right\} $
endowed with the multiplication rule \ $\lambda _{\alpha }\lambda _{\beta
}=\lambda _{\alpha +\beta }$ if \ $\alpha +\beta \leq 3;$ $\lambda _{\alpha
}\lambda _{\beta }=\lambda _{\alpha +\beta -4}$ if \ $\alpha +\beta >3$,
lead to the AdS$\mathcal{L}_{5}$-algebra, whose generators satisfy the
following commutation relations%
\begin{align}
\left[ J_{ab},J_{cd}\right] & =\eta _{bc}J_{ad}+\eta _{ad}J_{bc}-\eta
_{ac}J_{bd}-\eta _{bd}J_{ac},  \notag \\
\left[ J_{ab},Z_{cd}\right] & =\eta _{bc}Z_{ad}+\eta _{ad}Z_{bc}-\eta
_{ac}Z_{bd}-\eta _{bd}Z_{ac},  \notag \\
\left[ Z_{ab},Z_{cd}\right] & =\eta _{bc}J_{ad}+\eta _{ad}J_{bc}-\eta
_{ac}J_{bd}-\eta _{bd}J_{ac},  \notag \\
\left[ J_{ab},P_{c}\right] & =\eta _{bc}P_{a}-\eta _{ac}P_{b},\text{ \ \ \ }%
\left[ P_{a},P_{b}\right] =Z_{ab},  \notag \\
\left[ J_{ab},Z_{c}\right] & =\eta _{bc}Z_{a}-\eta _{ac}Z_{b},\text{ \ \ \ }%
\left[ P_{a},Z_{b}\right] =J_{ab},  \notag \\
\left[ Z_{ab},P_{c}\right] & =\eta _{bc}Z_{a}-\eta _{ac}Z_{b},\text{ \ \ \ }%
\left[ Z_{a},Z_{b}\right] =Z_{ab},  \notag \\
\left[ Z_{ab},Z_{c}\right] & =\eta _{bc}P_{a}-\eta _{ac}P_{b}.  \label{exp2}
\end{align}

\textbf{AdS}$\mathcal{L}_{6}$-\textbf{algebra: }This algebra is obtained
using the semigroup $S_{\mathcal{M}}^{\left( 4\right) }=\left\{ \lambda
_{0},\lambda _{1},\lambda _{2},\lambda _{3},\lambda _{4}\right\} $ endowed
with the multiplication rule $\lambda _{\alpha }\lambda _{\beta }=\lambda
_{\alpha +\beta }$ if \ $\alpha +\beta \leq 4;$ $\lambda _{\alpha }\lambda
_{\beta }=\lambda _{\alpha +\beta -4}$ if \ $\alpha +\beta >4$. Their
generators satisfy the following commutation relations 
\begin{align}
\left[ J_{ab},J_{cd}\right] & =\eta _{bc}J_{ad}+\eta _{ad}J_{bc}-\eta
_{ac}J_{bd}-\eta _{bd}J_{ac},  \notag \\
\left[ J_{ab},Z_{cd}^{\left( 1\right) }\right] & =\eta _{bc}Z_{ad}^{\left(
1\right) }+\eta _{ad}Z_{bc}^{\left( 1\right) }-\eta _{ac}Z_{bd}^{\left(
1\right) }-\eta _{bd}Z_{ac}^{\left( 1\right) },  \notag \\
\left[ J_{ab},Z_{cd}^{\left( 2\right) }\right] & =\eta _{bc}Z_{ad}^{\left(
2\right) }+\eta _{ad}Z_{bc}^{\left( 2\right) }-\eta _{ac}Z_{bd}^{\left(
2\right) }-\eta _{bd}Z_{ac}^{\left( 2\right) },  \notag \\
\left[ Z_{ab}^{\left( 1\right) },Z_{cd}^{\left( 1\right) }\right] & =\eta
_{bc}Z_{ad}^{\left( 2\right) }+\eta _{ad}Z_{bc}^{\left( 2\right) }-\eta
_{ac}Z_{bd}^{\left( 2\right) }-\eta _{bd}Z_{ac}^{\left( 2\right) },  \notag
\\
\left[ Z_{ab}^{\left( 1\right) },Z_{cd}^{\left( 2\right) }\right] & =\eta
_{bc}Z_{ad}^{\left( 1\right) }+\eta _{ad}Z_{bc}^{\left( 1\right) }-\eta
_{ac}Z_{bd}^{\left( 1\right) }-\eta _{bd}Z_{ac}^{\left( 1\right) },  \notag
\\
\left[ Z_{ab}^{\left( 2\right) },Z_{cd}^{\left( 2\right) }\right] & =\eta
_{bc}Z_{ad}^{\left( 2\right) }+\eta _{ad}Z_{bc}^{\left( 2\right) }-\eta
_{ac}Z_{bd}^{\left( 2\right) }-\eta _{bd}Z_{ac}^{\left( 2\right) },  \notag
\\
\left[ J_{ab},P_{c}\right] & =\eta _{bc}P_{a}-\eta _{ac}P_{b},\text{ \ \ \ }%
\left[ J_{ab},Z_{c}\right] =\eta _{bc}Z_{a}-\eta _{ac}Z_{b},  \notag \\
\left[ Z_{ab}^{\left( 1\right) },P_{c}\right] & =\eta _{bc}Z_{a}-\eta
_{ac}Z_{b},\text{ \ \ \ }\left[ Z_{ab}^{\left( 1\right) },Z_{c}\right] =\eta
_{bc}P_{a}-\eta _{ac}P_{b},  \notag \\
\left[ Z_{ab}^{\left( 2\right) },P_{c}\right] & =\eta _{bc}P_{a}-\eta
_{ac}P_{b},\text{ \ \ \ }\left[ Z_{ab}^{\left( 2\right) },Z_{c}\right] =\eta
_{bc}Z_{a}-\eta _{ac}Z_{b},  \notag \\
\left[ P_{a},P_{b}\right] & =Z_{ab}^{\left( 1\right) },\text{ \ \ \ }\left[
P_{a},Z_{b}\right] =Z_{ab}^{\left( 2\right) },\text{ \ \ \ }\left[
Z_{a},Z_{b}\right] =Z_{ab}^{\left( 1\right) }.  \label{exp3}
\end{align}

\section{$\mathfrak{B}_{4}$, $\mathfrak{B}_{5}$ and $\mathfrak{B}_{6}$
algebras}

$\mathfrak{B}_{4}$-\textbf{algebra}: This algebra normally called Maxwell
algebra can be obtained from AdS$\mathcal{L}_{4}$ by means of In\"{o}n\"{u}%
--Wigner contraction. \ In fact, rescaling $P_{a}\rightarrow \lambda P_{a}$, 
$Z_{ab}\rightarrow \lambda ^{2}Z_{ab}$ in (\ref{exp1}) and then taking the
limit $\lambda \rightarrow \infty $ we obtain the Maxwell algebra. \ 

$\mathfrak{B}_{5}$-\textbf{algebra:} This algebra can be obtained from AdS$%
\mathcal{L}_{5}$ algebra rescaling $P_{a}\rightarrow \lambda P_{a}$, $%
Z_{ab}\rightarrow \lambda ^{2}Z_{ab},$ $Z_{a}\rightarrow \lambda ^{3}Z_{a}$
in Eq. \ref{exp2} and then taking the limit $\lambda \rightarrow \infty $.

$\mathfrak{B}_{5}$-\textbf{algebra: }This algebra can also obtained from AdS$%
\mathcal{L}_{6}$ by means of rescaling $P_{a}\rightarrow \lambda P_{a}$, $%
Z_{a}\rightarrow \lambda ^{3}Z_{a}$, $Z_{ab}^{(1)}\rightarrow \lambda
^{2}Z_{ab}^{(1)},$ $Z_{ab}^{\left( 2\right) }\rightarrow \lambda
^{4}Z_{ab}^{\left( 2\right) }$ in (\ref{exp3}) and then taking the limit $%
\lambda \rightarrow \infty $

\subsection{\textbf{Semidirect sum structure of the} $\mathfrak{B}_{4}$, $%
\mathfrak{B}_{5}$ \textbf{and} $\mathfrak{B}_{6}$ \textbf{algebras}}

\textbf{Poincar\'{e} \'{a}lgebra: }the generators of the Poincar\'{e}
algebra are the generator of the translation group $T^{4}$ and the generator
of the Lorentz rotation group $so(3,1)$: $\left( P_{a},J_{ab}\right) $ which
satisfy the following commutation relations

\begin{align}
\left[ J_{ab},J_{cd}\right] & =\eta _{bc}J_{ad}+\eta _{ad}J_{bc}-\eta
_{ac}J_{bd}-\eta _{bd}J_{ac},  \notag \\
\left[ J_{ab},P_{c}\right] & =\eta _{bc}P_{a}-\eta _{ac}P_{b},  \notag \\
\left[ P_{a},P_{b}\right] & =0.  \label{ej1}
\end{align}

From (\ref{ej1}) we see that $T^{4}$ is an ideal of the Poincar\'{e} algebra
since $\left[ T^{4},T^{4}\right] \subset T^{4}$ and $\left[ so(3,1),T^{4}%
\right] \subset T^{4}$, which means that $iso(3,1)=so(3,1)\uplus T^{4}$.

\textbf{\ Maxwell algebra:} the generators of Maxwell algebra $\left(
P_{a},J_{ab},Z_{ab}\right) $ satisfy the commutation relations

\begin{align}
\left[ J_{ab},J_{cd}\right] & =\eta _{bc}J_{ad}+\eta _{ad}J_{bc}-\eta
_{ac}J_{bd}-\eta _{bd}J_{ac},  \notag \\
\left[ J_{ab},P_{c}\right] & =\eta _{bc}P_{a}-\eta _{ac}P_{b},\text{ \ \ \ }%
\left[ P_{a},P_{b}\right] =\Lambda Z_{ab},  \notag \\
\left[ J_{ab},Z_{cd}\right] & =\eta _{bc}Z_{ad}+\eta _{ad}Z_{bc}-\eta
_{ac}Z_{bd}-\eta _{bd}Z_{ac},  \notag \\
\left[ Z_{ab},Z_{cd}\right] & =0,\text{ \ }\left[ Z_{ab},P_{c}\right] =0,
\label{ej2}
\end{align}%
from where we see that the subset of generators $\mathcal{M}^{I}=\left(
P_{a},Z_{ab}\right) $ is an ideal of Maxwell's algebra since $\left[ 
\mathcal{M}^{I},\mathcal{M}^{I}\right] \subset \mathcal{M}^{I}$, $\left[
so(3,1),\mathcal{M}^{I}\right] \subset \mathcal{M}^{I}$. This means that the
Maxwell algebra $\mathcal{M}$ is the semidirect sum of the Lorentz algebra $%
so(3,1)$ and the ideal $\mathcal{M}^{I}$, that is $\mathcal{M}=so(3,1)\uplus 
\mathcal{M}^{I}$.

\textbf{Generalized} \textbf{Poincar\'{e} algebra} $\mathfrak{B}_{5}$: the
generators of this algebra $\left( P_{a},J_{ab},Z_{ab},Z_{a}\right) $
satisfy the commutation relations

\begin{align}
\left[ P_{a},P_{b}\right] & =\Lambda Z_{ab},  \notag \\
\left[ J_{ab},P_{c}\right] & =\eta _{bc}P_{a}-\eta _{ac}P_{b},  \notag \\
\left[ J_{ab},J_{cd}\right] & =\eta _{bc}J_{ad}+\eta _{ad}J_{bc}-\eta
_{ac}J_{bd}-\eta _{bd}J_{ac},  \notag \\
\left[ J_{ab},Z_{cd}\right] & =\eta _{bc}Z_{ad}+\eta _{ad}Z_{bc}-\eta
_{ac}Z_{bd}-\eta _{bd}Z_{ac},  \notag \\
\left[ J_{ab},Z_{c}\right] & =\eta _{bc}Z_{a}-\eta _{ac}Z_{b},  \notag \\
\left[ Z_{ab},P_{c}\right] & =\eta _{bc}Z_{a}-\eta _{ac}Z_{b},  \notag \\
\left[ Z_{ab},Z_{c}\right] & =0,\text{ \ \ }\left[ Z_{ab},Z_{cd}\right] =0,%
\text{ \ \ \ }  \notag \\
\text{ \ \ \ }\left[ P_{a},Z_{b}\right] & =0,\text{ \ }\left[ Z_{a},Z_{b}%
\right] =0,  \label{ej3}
\end{align}%
from where we see that the subset of generators $\mathfrak{B}_{5}^{I}=\left(
P_{a},Z_{ab},Z_{a}\right) $ is an ideal of $\mathfrak{B}_{5}$ algebra since $%
\left[ \mathfrak{B}_{5}^{I},\mathfrak{B}_{5}^{I}\right] \subset \mathfrak{B}%
_{5}^{I}$, $\left[ so(3,1),\mathfrak{B}_{5}^{I}\right] \subset \mathfrak{B}%
_{5}^{I}$. This means that the Maxwell algebra $\mathfrak{B}_{5}$ is the
semidirect sum of the Lorentz algebra $so(3,1)$ and the ideal $\mathfrak{B}%
_{5}^{I}$, that is $\mathfrak{B}_{5}=so(3,1)\uplus \mathfrak{B}_{5}^{I}$.

\textbf{Generalized} \textbf{Poincar\'{e} algebra} $\mathfrak{B}_{6}$: the
generators of this algebra $\left( P_{a},J_{ab},Z_{cd}^{\left( 1\right)
},Z_{ab}^{\left( 2\right) },Z_{a}\right) $ satisfy the commutation relations

\begin{align}
\left[ J_{ab},J_{cd}\right] & =\eta _{bc}J_{ad}+\eta _{ad}J_{bc}-\eta
_{ac}J_{bd}-\eta _{bd}J_{ac},  \notag \\
\left[ J_{ab},Z_{cd}^{\left( 1\right) }\right] & =\eta _{bc}Z_{ad}^{\left(
1\right) }+\eta _{ad}Z_{bc}^{\left( 1\right) }-\eta _{ac}Z_{bd}^{\left(
1\right) }-\eta _{bd}Z_{ac}^{\left( 1\right) },  \notag \\
\left[ J_{ab},Z_{cd}^{\left( 2\right) }\right] & =\eta _{bc}Z_{ad}^{\left(
2\right) }+\eta _{ad}Z_{bc}^{\left( 2\right) }-\eta _{ac}Z_{bd}^{\left(
2\right) }-\eta _{bd}Z_{ac}^{\left( 2\right) },  \notag \\
\left[ J_{ab},P_{c}\right] & =\eta _{bc}P_{a}-\eta _{ac}P_{b},\text{ \ \ \ }%
\left[ J_{ab},Z_{c}\right] =\eta _{bc}Z_{a}-\eta _{ac}Z_{b},  \notag \\
\left[ Z_{ab}^{\left( 1\right) },Z_{cd}^{\left( 1\right) }\right] & =\eta
_{bc}Z_{ad}^{\left( 2\right) }+\eta _{ad}Z_{bc}^{\left( 2\right) }-\eta
_{ac}Z_{bd}^{\left( 2\right) }-\eta _{bd}Z_{ac}^{\left( 2\right) },  \notag
\\
\left[ Z_{ab}^{\left( 1\right) },P_{c}\right] & =\eta _{bc}Z_{a}-\eta
_{ac}Z_{b},\text{ \ }  \notag \\
\left[ P_{a},P_{b}\right] & =\Lambda Z_{ab}^{\left( 1\right) },\text{ \ \ }%
\left[ P_{a},Z_{b}\right] =Z_{ab}^{\left( 2\right) },  \notag \\
\left[ Z_{ab}^{\left( 1\right) },Z_{c}\right] & =0,\text{ \ }\left[
Z_{ab}^{\left( 1\right) },Z_{cd}^{\left( 2\right) }\right] =0,  \notag \\
\left[ Z_{ab}^{\left( 2\right) },Z_{cd}^{\left( 2\right) }\right] & =0,\text{
\ }\left[ Z_{ab}^{\left( 2\right) },P_{c}\right] =0  \notag \\
\text{\ }\left[ Z_{ab}^{\left( 2\right) },Z_{c}\right] & =0,\text{ \ }\left[
Z_{a},Z_{b}\right] =0,  \label{ej4'}
\end{align}

from where we see that the subset of generators $\mathfrak{B}_{6}^{I}=\left(
P_{a},Z_{cd}^{\left( 1\right) },Z_{ab}^{\left( 2\right) },Z_{a}\right) $ is
an ideal of $\mathfrak{B}_{6}$ algebra since $\left[ \mathfrak{B}_{6}^{I},%
\mathfrak{B}_{6}^{I}\right] \subset \mathfrak{B}_{6}^{I}$, $\left[ so(3,1),%
\mathfrak{B}_{6}^{I}\right] \subset \mathfrak{B}_{6}^{I}$. This means that
the Maxwell algebra $\mathfrak{B}_{6}$ is the semidirect sum of the Lorentz
algebra $so(3,1)$ and the ideal $\mathfrak{B}_{6}^{I}$, that is $\mathfrak{B}%
_{6}=so(3,1)\uplus \mathfrak{B}_{6}^{I}$.

\section{\textbf{Appendix B: About }$AdS\mathcal{L}_{6}$\textbf{-gravity}}

In this case, the $AdS\mathcal{L}_{6}$-valued one-form gauge connetion is
given by

\begin{equation*}
A=\frac{1}{2}\omega ^{ab}J_{ab}+\frac{1}{l}e^{a}P_{a}+\frac{1}{2}%
k^{ab}Z_{ab}+\frac{1}{l}h^{a}Z_{a}+\frac{1}{2}q^{ab}\hat{Z}_{ab}
\end{equation*}%
where now the generators $P_{a},J_{ab},Z_{ab},Z_{a},\hat{Z}_{ab}$ satisfy
the $AdS\mathcal{L}_{6}$\textbf{-}algebra commutation relations (\ref{exp3}%
). \ Following the same procedure of the previous sections we find that the
associated two-form curvature are given by%
\begin{equation*}
F=\frac{1}{2}R^{ab}J_{ab}+\frac{1}{l}\mathcal{\bar{T}}^{\text{ }a}P_{a}+%
\frac{1}{2}\bar{F}^{ab}Z_{ab}+\frac{1}{l}\bar{H}^{a}Z_{a}+\frac{1}{2}Q^{ab}%
\hat{Z}_{ab},
\end{equation*}%
with%
\begin{align}
\mathcal{\bar{T}}^{\text{ }a}& =T^{a}+q_{\;b}^{a}e^{b}+k_{\;b}^{a}h^{b}, \\
R^{ab}& =d\omega ^{ab}+\omega _{\text{ \ }c}^{a}\omega ^{cb}, \\
\bar{F}^{ab}& =Dk^{ab}+q_{\;\;c}^{[a}k^{c|b]}+\Lambda e^{a}e^{b}+\Lambda
h^{a}h^{b}, \\
Q^{ab}& =Dq^{ab}+k_{\text{ \ }c}^{a}k^{cb}+q_{\;c}^{a}q^{cb}+\Lambda
e^{[a}h^{b]}, \\
\bar{H}^{a}& =Dh^{a}+k_{\;c}^{a}e^{c}+q_{\;c}^{a}h^{c},
\end{align}%
where $T^{a}$ and $R^{ab}$ are the standard torsion and curvature, $\bar{F}%
^{ab}$ is the curvature associated to the generator $Z_{ab}$ and $Q^{ab},%
\bar{H}^{a}$ are the curvature associated to the generators $\hat{Z}_{ab}$
and $Z_{a}$ respectively.

The covariant derivative of the curvatures lead to the following Bianchi
identities%
\begin{equation*}
DF=\frac{1}{2}DR^{ab}J_{ab}+\frac{1}{l}D\mathcal{\bar{T}}^{\text{ }a}P_{a}+%
\frac{1}{2}D\bar{F}^{ab}Z_{ab}+\frac{1}{l}D\bar{H}^{a}Z_{a}+\frac{1}{2}%
DQ^{ab}\hat{Z}_{ab},
\end{equation*}%
with%
\begin{align}
dR^{ab}+\omega _{\;c}^{a}R^{cb}-\omega _{\;c}^{b}R^{ca} =DR^{ab}&=0, \\
D\mathcal{\bar{T}}^{a}+R_{\;b}^{a}e^{b}+q_{\;b}^{a}\mathcal{\tilde{T}}%
^{b}+Q_{\;b}^{a}e^{b} + k^{a}_{\;c} \bar{H}^{c} & =0, \\
D\bar{F}^{ab}+k_{\;\;c}^{[a}R^{c|b]}+q_{\;\;c}^{[a}\bar{F}%
^{c|b]}+k_{\;\;c}^{[a}Q^{c|b]}+\frac{2}{l^{2}}e^{a}\mathcal{\bar{T}}^{b}+%
\frac{2}{l^{2}}h^{a}\bar{H}^{b}& =0, \\
D\bar{H}^{a}+R_{\text{ \ }b}^{a}h^{b}+k_{\;b}^{a}\mathcal{\bar{T}}^{\text{ }%
b}+\bar{F}_{\;b}^{a}e^{b}+q_{\;b}^{a}\bar{H}^{b}+Q_{\text{ \ }b}^{a}h^{b}&
=0, \\
DQ^{ab}+q_{\;\;c}^{[a}R^{c|b]}+k_{\;\;c}^{[a}\bar{F}^{c|b]}+q_{\;%
\;c}^{[a}Q^{c|b]}+\frac{2}{l^{2}}h^{a}\mathcal{\bar{T}}^{b} + \frac{2}{l^{2}}
e^{a} \bar{H}^{b}& =0
\end{align}

The gauge potencials transform as 
\begin{equation}
\delta _{\varepsilon }A=D\varepsilon =d\varepsilon +\left[ A,\varepsilon %
\right] ,
\end{equation}%
where $\varepsilon $ is a $AdS\mathcal{L}_{6}$ algebra valued parameter
given by 
\begin{equation}
\varepsilon =\frac{1}{2}\pi ^{ab}J_{ab}+\frac{1}{l}\rho ^{a}P_{a}+\frac{1}{2}%
\xi ^{ab}Z_{ab}+\frac{1}{l}\sigma ^{a}Z_{a}+\frac{1}{2}\chi ^{ab}\hat{Z}%
_{ab},  \notag
\end{equation}

\bigskip Using the commutation relation of the $AdS\mathcal{L}_{6}$\textbf{-}%
algebra we find

\begin{align}
\delta \omega ^{ab}& =d\pi ^{ab}+\omega _{\;\;c}^{[a}\pi ^{c|b]}, \\
\delta e^{a}& =d\rho ^{a}+\pi _{\;c}^{a}e^{c}+\omega _{\;c}^{a}\rho
^{c}+q_{\;c}^{a}\rho ^{c}+\xi _{\;c}^{a}h^{c}+k_{\;c}^{a}\sigma ^{c}+\chi
_{\;c}^{a}e^{c}, \\
\delta k^{ab}& =d\xi ^{ab}+k_{\;\;c}^{[a}\pi ^{c|b]}+\omega _{\;\;c}^{[a}\xi
^{c|b]}+q_{\;\;c}^{[a}\xi ^{c|b]}+k_{\;\;c}^{[a}\chi ^{c|b]}+\Lambda
e^{[a}\rho ^{b]}+\Lambda h^{[a}\sigma ^{b]}, \\
\delta h^{a}& =d\sigma ^{a}+\pi _{\;c}^{a}h^{c}+k_{\;c}^{a}\rho ^{c}+\xi
_{\;c}^{a}e^{c}+\omega _{\;c}^{a}\sigma ^{c}+q_{\;c}^{a}\sigma ^{c}+\chi
_{\;c}^{a}h^{c}, \\
\delta q^{ab}& =d\chi ^{ab}+q_{\;\;c}^{[a}\pi ^{c|b]}+k_{\;\;c}^{[a}\xi
^{c|b]}+\omega _{\;\;c}^{[a}\chi ^{c|b]}+q_{\;\;c}^{[a}\chi ^{c|b]}+\Lambda
h^{[a}\rho ^{b]}+\Lambda e^{[a}\sigma ^{b]}.
\end{align}

On the another hand, from $\delta _{\varepsilon }F=\left[ F,\varepsilon %
\right] $ we find that the curvatures transform as

\begin{align}
\delta R^{ab}& =R_{\;\;c}^{[a}\pi ^{c|b]}, \\
\delta \mathcal{\bar{T}}^{\text{ }a}& =\pi _{\;c}^{a}\mathcal{\bar{T}}^{%
\text{ }c}+R_{\;c}^{a}\rho ^{c}+Q_{\;c}^{a}\rho ^{c}+\xi _{\;c}^{a}\bar{H}%
^{c}+\bar{F}_{\;c}^{a}\sigma ^{c}+\chi _{\;c}^{a}\mathcal{\bar{T}}^{\text{ }%
c}, \\
\delta \bar{F}^{ab}& =\bar{F}_{\;\;c}^{[a}\pi ^{c|b]}+R_{\;\;c}^{[a}\xi
^{c|b]}+Q_{\;\;c}^{[a}\xi ^{c|b]}+\bar{F}_{\;\;c}^{[a}\chi ^{c|b]}+\Lambda 
\mathcal{\bar{T}}^{[a}\rho ^{b]}+\Lambda \bar{H}^{[a}\sigma ^{b]}, \\
\delta \bar{H}^{a}& =\pi _{\;c}^{a}\bar{H}^{c}+\bar{F}_{\;c}^{a}\rho
^{c}+\xi _{\;c}^{a}\mathcal{\bar{T}}^{c}+R_{\;c}^{a}\sigma
^{c}+Q_{\;c}^{a}\sigma ^{c}+\chi _{\;c}^{a}\bar{H}^{c}, \\
\delta Q^{ab}& =Q_{\;\;c}^{[a}\pi ^{c|b]}+\bar{F}_{\;\;c}^{[a}\xi
^{c|b]}+R_{\;\;c}^{[a}\chi ^{c|b]}+Q_{\;\;c}^{[a}\chi ^{c|b]}+\Lambda \bar{H}%
^{[a}\rho ^{b]}+\Lambda \mathcal{\bar{T}}^{[a}\sigma ^{b]}.
\end{align}

\subsection{\textbf{Field Equations for }$AdSL_{6}$\textbf{-gravity}}

The same procedure of the previous section leads to an action for the $AdS%
\mathcal{L}_{6}$-gravity constructed from the following 4-forms invariant

\begin{align}
-\frac{1}{2\kappa\Lambda }\mathcal{L}_{1}&=-\frac{1}{2\kappa\Lambda }%
\varepsilon _{abcd}R^{ab} \bar{F}^{cd}  \notag \\
&=\mathcal{L}_{E-H}-\frac{1}{2\kappa}\varepsilon _{abcd}R^{ab}h^{c}h^{d} -%
\frac{1}{2\kappa\Lambda }\varepsilon _{abcd}R^{ab}q_{\;\;e}^{[c}k^{e|d]}
\end{align}

\begin{align}
\frac{\lambda }{2\kappa\Lambda ^{2}}\mathcal{L}_{2}&=\frac{\lambda }{%
4\kappa\Lambda ^{2}}\varepsilon _{abcd}\bar{F}^{ab} \bar{F}^{cd}  \notag \\
&=\mathcal{L}_{cosm}+\frac{\lambda }{4\kappa\Lambda ^{2}}\varepsilon
_{abcd}\left\{ Dk^{ab}Dk^{cd}+2Dk^{ab}q_{\;\;e}^{[c}k^{e|d]} \right.  \notag
\\
&+2\Lambda Dk^{ab}e^{c}e^{d} +2\Lambda Dk^{ab}h^{c}h^{d}+2\Lambda
q_{\;\;e}^{[a}k^{e|b]}e^{c}e^{d}+2\Lambda q_{\;\;e}^{[a}k^{e|b]}h^{c}h^{d} 
\notag \\
& \left.+2\Lambda ^{2}e^{a}e^{b}h^{c}h^{d}
+q_{\;\;e}^{[a}k^{e|b]}q_{\;\;f}^{[c}k^{f|d]}+\Lambda
^{2}h^{a}h^{b}h^{c}h^{d}\right\} ,
\end{align}

Defining $\mu =\lambda /\Lambda $ we propose%
\begin{equation*}
\mathcal{L}_{AdS\mathcal{L}_{6}}=-\frac{\mu }{2\kappa\Lambda }\mathcal{L}%
_{1}+\frac{\mu ^{2}}{2\kappa\Lambda }\mathcal{L}_{2},
\end{equation*}
\begin{align}
\mathcal{L}& =\mathcal{L}_{E-H}+\mathcal{L}_{cosm}-\frac{1}{2\kappa }%
\varepsilon _{abcd}R^{ab}h^{c}h^{d}-\frac{1}{2\kappa \Lambda }\varepsilon
_{abcd}R^{ab}q_{\;\;e}^{[c}k^{e|d]}  \notag \\
& +\frac{\mu ^{2}}{4\kappa \lambda }\varepsilon _{abcd}Dk^{ab}Dk^{cd}+\frac{%
\mu ^{2}}{2\kappa \lambda }\varepsilon _{abcd}Dk^{ab}q_{\;\;e}^{[c}k^{e|d]} 
\notag \\
& +\frac{\mu }{2\kappa }\varepsilon _{abcd}Dk^{ab}e^{c}e^{d}+\frac{\mu }{%
2\kappa }\varepsilon _{abcd}Dk^{ab}h^{c}h^{d}  \notag \\
& +\frac{\mu }{2\kappa }\varepsilon _{abcd}q_{\;\;e}^{[a}k^{e|b]}e^{c}e^{d}+%
\frac{\mu }{2\kappa }\varepsilon _{abcd}q_{\;\;e}^{[a}k^{e|b]}h^{c}h^{d} 
\notag \\
& +\frac{\lambda }{2\kappa }\varepsilon _{abcd}e^{a}e^{b}h^{c}h^{d}+\frac{%
\lambda }{4\kappa \Lambda ^{2}}q_{\;\;e}^{[a}k^{e|b]}q_{\;\;e}^{[c}k^{e|d]}+%
\frac{\lambda }{4\kappa }\varepsilon _{abcd}h^{a}h^{b}h^{c}h^{d},
\end{align}

The variation of $\mathcal{L}_{AdS\mathcal{L}_{6}}$ with respect to $e^{a},$ 
$h^{a},$ $\omega ^{ab},$ $k^{ab},$ and $q^{ab}$ leads to the following
equations of motion

\begin{align}
\left[ L\right] _{e^{a}}& =\frac{1}{\kappa }\varepsilon _{abcd}\left( \mu
\left( Dk^{ab}+q_{\;\;e}^{[a}k^{e|b]}+\Lambda h^{a}h^{b}+\Lambda
e^{a}e^{b}\right) e^{c}-R^{ab}e^{c}\right) =0 \\
& =\frac{1}{\kappa }\varepsilon _{abcd}\left( \mu \bar{F}%
^{ab}e^{c}-R^{ab}e^{c}\right) =0
\end{align}

\begin{align}
\left[ L\right] _{h^{a}}& =\frac{1}{\kappa }\varepsilon _{abcd}\left(
-R^{ab}h^{c}+\mu \left\{ Dk^{ab}+q_{\;\;e}^{[a}k^{e|b]}+\Lambda
e^{a}e^{b}+\Lambda h^{a}h^{b}\right\} h^{c}\right) =0 \\
& =\frac{1}{\kappa }\varepsilon _{abcd}\left( -R^{ab}h^{c}+\mu \bar{F}%
^{ab}h^{c}\right) =0
\end{align}

\begin{equation*}
\left[ L\right] _{\omega ^{ab}}=\frac{1}{\kappa}\varepsilon_{abcd} \left\{ -%
\frac{\mu^{2}}{\lambda}k^{a}_{\;\;e}\Tilde{F}^{eb} +\frac{1}{\Lambda}
D\left( q^{[a}_{\;\;e}k^{e|b]} \right) - T^{a}e^{b}-Dh^{a}h^{b}\right\}=0
\end{equation*}

\begin{equation*}
\left[ L\right] _{k^{ab}}=\frac{1}{\kappa }\varepsilon _{abcd}\left( \mu
T^{a}e^{b}+\mu Dh^{a}h^{b}+\frac{\mu ^{2}}{2\lambda }R_{\,\,\,e}^{a}k^{eb}+%
\frac{\mu ^{2}}{\lambda }q_{\,\,\,e}^{a}\tilde{F}^{eb}+\frac{\mu ^{2}}{%
\lambda }D\left( q_{\,\,\,e}^{a}k^{eb}\right) \right) =0
\end{equation*}

\begin{equation*}
\left[ L\right] _{q^{ab}}=\frac{1}{\kappa }\varepsilon
_{abcd}k_{\;\;e}^{a}\left\{ R^{eb}-\frac{\mu ^{2}}{\lambda }\Tilde{F}%
^{eb}\right\} =0.
\end{equation*}

\end{document}